\documentclass[a4paper]{article}
\usepackage{jheppub}

\usepackage{lscape}
\usepackage{amsfonts}
\usepackage{mathrsfs}

\def\tr{{\rm Tr }}

\def\bea{\begin{eqnarray}}
\def\eea{\end{eqnarray}}
\def\nn{\nonumber}

\def\lmatrix{\left(\begin{array}}
\def\rmatrix{\end{array}\right)}
\def\gsim{\mathrel{\rlap{\lower4pt\hbox{\hskip1pt$\sim$}}\raise1pt\hbox{$>$}}}
\def\lsim{\mathrel{\rlap{\lower4pt\hbox{\hskip1pt$\sim$}}\raise1pt\hbox{$<$}}}
\def\bi{\begin{itemize}}
\def\ei{\end{itemize}}
\def\msbar{\overline{\rm MS\kern-0.5pt}\kern0.5pt}
\def\rho{\varrho}

\title{Vector fields, RG flows and emergent gauge symmetry}

\author{Daniel Nogradi}

\affiliation{Eotvos University, Department of Theoretical Physics \\ Pazmany Peter setany 1/a, Budapest 1117, Hungary}

\emailAdd{nogradi@bodri.elte.hu}

\abstract{We consider the most general perturbatively renormalizable theory of vector fields in four 
dimensions with a global $SU(N)$ symmetry and massless couplings. The
Lagrangian contains 1 quadratic, 2 cubic and 4 quartic couplings. The RG flow among this set
of 7 couplings is computed to 1-loop and a rich phase diagram is mapped out; in particular it is shown that 
a finite number of asymptotically free RG-flows exist corresponding to non-trivial fixed points for the ratios of the
couplings. None of these are gauge theories, i.e. possess only global $SU(N)$ invariance but not
a local one. We also include the most general ghost couplings, still with global $SU(N)$
invariance, and compute the RG flow to 1-loop for all 9 resulting couplings. Again asymptotically free RG flows
exist with non-trivial fixed points for the ratios of couplings. It is shown that Yang-Mills theory emerges at a
particular fixed point. The theories at the other fixed points are marginally relevant gauge symmetry violating
perturbations of Yang-Mills theory. The large-$N$ limit is also investigated in detail.
}

\begin{document}

\maketitle

\section{Introduction}
\label{introduction}

All interactions in the Standard Model are carried by spin-1 fields (vector fields). As long as they are massless
only 2 of the 4 degrees of freedom are physical,
due to gauge invariance which guarantees a unitary Minkowskian quantum field theory. If the gauge
group is non-abelian these vector fields interact with themselves which in turn leads to asymptotic freedom. The
form of the self-interaction is constrained by gauge invariance and many terms in the Lagrangian which would otherwise
be allowed by locality, Lorentz invariance and perturbative renormalizability, are forbidden.

In this work we investigate what happens with the renormalization group (RG) flows and asymptotic freedom if gauge
invariance is relaxed and the forbidden terms are allowed to be present. Symmetry under {\em global} transformations
is still imposed. But otherwise the most general local, Lorentz invariant and perturbatively renormalizable Lagrangian 
is considered for a vector field $A_\mu^a(x)$ in the adjoint of $SU(N)$ with dimensionless couplings. 
In total, there are 7 non-trivial couplings,
as opposed to one in Yang-Mills theory.

We will see that the cubic and quartic interactions allow for asymptotically free RG flows with non-trivial UV fixed points
for the ratios of couplings. These ratios correspond to particular directions in which the Gaussian fixed point is
approached towards the UV. The resulting non-trivial quantum field theories are not gauge theories (and hence
are non-unitary if continued to Minkowskian signature) but 
are nonetheless perfectly well-defined and local in Euclidean signature.
The constraint $\partial_\mu A_\mu^a =
0$ naturally arises for $N > 5$ and also in the large-$N$ limit. In this case, for $N>5$, the main result is that there
exists a unique asymptotically free Euclidean quantum field theory with an action bounded from below and stable 
RG fixed point in the UV for the ratios of couplings. Generally, a rich phase diagram is mapped out for all 
$N$ including only partially stable RG fixed points for the ratios.

In order to embed perturbative Yang-Mills theory in our setup the most general couplings with
ghost fields are also considered, bringing the number of couplings to 9. Global $SU(N)$ invariance is still
assumed. Again asymptotically free RG flows
are found with non-trivial UV fixed points for the ratios and Yang-Mills theory is identified as one of the fixed points in the
space of ratios. At this particular fixed point gauge symmetry is an emergent phenomenon since at every other point only
a {\em global} symmetry is present. Unitarity is emergent as well since only at the particular fixed point corresponding to
Yang-Mills theory does the theory admit a unitary Minkowskian version.

In the large-$N$ limit we also study in more detail the vicinity of gauge theory in the space of all couplings and
identify several marginal perturbations. With respect to some of these Yang-Mills theory is RG stable but there
are also directions with respect to which it is unstable.

Similar questions as the ones addressed in this work were discussed with $U(1)$ at the 1-loop level in 
\cite{Iliopoulos:1980zd} and rather qualitatively for the non-abelian case in
\cite{Forster:1980dg}.

The primary motivation for the present work is the desire to understand to what extent asymptotic freedom is special to
gauge theories in 4 dimensions. The main result is that gauge invariance is not the essential ingredient but rather the
cubic interaction. Without cubic interactions of the vector fields asymptotic freedom is not possible. But certain
cubic interactions do lead to asymptotic freedom even without gauge invariance.

The secondary motivation for the present work is an analogy with diffeomorphism 
breaking approaches to quantum gravity; see \cite{Catterall:2009nz} and references therein. In these
approaches the discretization breaks diffeomorphisms and it is hoped
that at particular corners of the coupling space it is recovered. What would be necessary for this scenario is
an emergent infinite dimensional symmetry group by tuning only a finite number of couplings, very much in analogy
with the results presented here, although with the obvious difference that quantum gravity is not expected to be an
asymptotically free theory.

The organization of the paper is as follows. In section \ref{lagrangianwithvectorfields} the most general Lagrangian is
spelled out with the desired properties and the 7 independent couplings are defined, followed by section
\ref{rgflowswithvectorfields} where the solutions for the 1-loop running are found and asymptotically free flows are
identified. Section \ref{rgflows} details the inclusion of ghosts fields and 2 new couplings accordingly. This section
contains our discussion about emergent gauge invariance and marginal perturbations of Yang-Mills theory.
Finally, in section \ref{conclusion} we close with an outlook to possible future studies.

\section{Lagrangian with vector fields}
\label{lagrangianwithvectorfields}

We will consider vector fields $A_\mu$ in the adjoint of $SU(N)$ and wish to construct the most general 
perturbatively renormalizable
Lagrangian in four dimensions with global $SU(N)$ and Lorentz invariance and massless couplings, 
i.e. classically scale invariant, but not conformal 
\cite{Jackiw:2011vz,ElShowk:2011gz,Nakayama:2013is},
theories. Dimensional regularization and the $\msbar$ scheme will be employed in Euclidean signature.

There are clearly 2 independent kinetic terms,
\bea
\tr \left( \partial_\mu A_\nu \partial_\mu A_\nu \right),\qquad\qquad \tr \left( \partial_\mu A_\mu \partial_\nu A_\nu
\right),
\label{a2}
\eea
there are also 2 independent cubic terms,
\bea
\tr \left( A_\nu A_\nu \partial_\mu A_\mu \right),\qquad\qquad \tr \left( A_\mu A_\nu \partial_\mu A_\nu \right),
\label{a3}
\eea
and 4 independent quartic terms,
\bea
( \tr A_\mu A_\mu )^2,\qquad \tr\left( A_\mu A_\nu\right) \tr\left( A_\mu A_\nu\right),\qquad \tr \left( A_\mu A_\mu
A_\nu A_\nu\right),\qquad \tr \left( A_\mu A_\nu A_\mu A_\nu \right),
\label{a4}
\eea
up to total derivatives. Terms higher than dimension 4 are not allowed by perturbative renormalizability. Once a
(Hermitian) basis
$T_a$ is chosen and the fields $A_\mu^a$ are canonically normalized with respect to the first kinetic term in (\ref{a2})
a possible parametrization of the most general Lagrangian in terms of 7 independent couplings is,
\bea
\label{lagrangian}
{\mathscr L} &=& \frac{1}{2} \partial_\mu A_\nu^a \partial_\mu A_\nu^a - 
\frac{1}{2}\left(1-\frac{1}{z}\right) (\partial_\mu A_\mu^a )^2 + 
h_1 {\tilde{\mathscr O}}_1 +  h_2 {\tilde{\mathscr O}}_2 + {\mathscr V}  \nn \\
{\tilde {\mathscr O}}_1 &=& A_\mu^a A_\nu^b \partial_\mu A_\nu^c d_{abc} \nn \\
{\tilde {\mathscr O}}_2 &=& A_\mu^a A_\nu^b \partial_\mu A_\nu^c f_{abc} \nn \\
{\mathscr V} &=& \sum_{i=1}^4 g_i {\mathscr O}_i \\
{\mathscr O}_1 &=& \frac{1}{8} A_\mu^a A_\mu^b A_\nu^c A_\nu^g d_{abe} d_{cge} = \tr(A_\mu A_\mu A_\nu A_\nu) - \frac{1}{N}\left(
\tr(A_\mu A_\mu) \right)^2 \geq 0 \nn \\
{\mathscr O}_2 &=& \frac{1}{8N}( A_\mu^a A_\mu^a )^2 = \frac{1}{2N} \left( \tr( A_\mu A_\mu ) \right)^2  \geq 0 \nn \\
{\mathscr O}_3 &=& \frac{1}{8N} A_\mu^a A_\mu^b A_\nu^a A_\nu^b = \frac{1}{2N} \tr(A_\mu A_\nu) \tr(A_\mu A_\nu) \geq 0 \nn \\
{\mathscr O}_4 &=& \frac{1}{4} A_\mu^a A_\mu^b A_\nu^c A_\nu^g f_{ace} f_{bge} = - \frac{1}{2} \tr[A_\mu,A_\nu]^2 \geq 0\;, \nn
\eea
where $f_{abc}$ are the structure constants of $SU(N)$ and $d_{abc}= 2\, \tr \left\{ T_a, T_b \right\} T_c$ 
is the totally symmetric tensor.
There is one non-trivial 2-point coupling, $z$, two 3-point couplings $(h_1,h_2)$ and four 4-point couplings
$(g_1,g_2,g_3,g_4)$. The explicit factors of $N$ included for $(g_2,g_3)$ are there for the sake of simple large-$N$
scaling relations. The couplings in the Lagrangian are bare couplings of course and when in later sections we discuss
RG flows and $\beta$-functions we will use the same notation for the renormalized couplings as well.

The special case of classical Yang-Mills theory with coupling $g$ (without gauge fixing terms) 
is given by the subspace $(z,g_1,g_2,g_3,g_4,h_1,h_2) = ( \infty, 0, 0, 0, g^2, 0, g)$.

It is easy to show that each ${\mathscr O}_i$ is indeed non-negative as shown in (\ref{lagrangian}). 
However it is a non-trivial question whether the full quartic potential ${\mathscr V}$ is 
positive semi-definite or not for given $g_i$. Unfortunately, 
we are not able to give a necessary and sufficient condition for $(g_1,g_2,g_3,g_4)$ leading to a positive semi-definite
${\mathscr V}$. One of the following two conditions is however necessary,
\bea
&g_1& \geq 0\;, \qquad g_2 + g_3 \geq - g_1 ( N - 2 ) \nn \\
&g_1& \leq 0\;, \qquad g_2 + g_3 \geq - g_1 \frac{2(N-2)^2}{N-1}\;,
\eea
which conditions can be obtained by studying the coefficient of $(A_\mu^a)^4$ for given $\mu$ and $a$. This coefficient
needs to be positive. At the same time any one of the following conditions is sufficient,
\bea
\begin{array}{cccc}
g_1 \geq 0\;, &\;\; 4 g_2 + g_3 \geq 0\;,     &\;\; g_3 \geq 0\;, &\;\; g_4 \geq 0 \\
g_1 \geq 0\;, &\;\; 4 g_2 + g_3 \geq 8 g_1\;, &\;\; g_3 \geq 0\;, &\;\; 3 g_4 \geq - 2 g_1 \\
g_1 \geq 0\;, &\;\; g_2 + g_3 \geq 0\;,       &\;\; g_3 \leq 0\;, &\;\; g_4 \geq 0 \\
g_1 \leq 0\;, &\;\; g_2 + 2(N-1)g_1 \geq 0\;, &\;\; g_3 \geq 0\;, &\;\; g_4 \geq 0 \\ 
\end{array}
\eea
as can be checked by completing squares in various ways. In any case it is not necessary to have $g_i \geq 0$ for all
couplings for the potential ${\mathscr V}$ to be non-negative.
Positivity of the full action also requires $z\geq0$.

Note that a positive semi-definite potential may not be strictly necessary for
well-defined Green's functions in quantum field theory, see for instance \cite{Fei:2014yja}, 
but clearly a positive potential places the construction on more solid footing.

There is a curious interplay between the couplings $z$ and $h_1$. The operator ${\tilde {\mathscr O}}_1$,
which corresponds to the coupling $h_1$, is a total derivative up to terms which are proportional 
to $\partial_\mu A_\mu^a$. On the other hand $z=0$ corresponds to the constraint $\partial_\mu A_\mu^a = 0$,
hence in this case ${\tilde{\mathscr O}}_1$ is a total derivative. This in turn means that if
$z=0$ the coupling $h_1$ can not appear at any order of the perturbative expansion, i.e. the dependence on $h_1$ can
only be through the product $z^n h_1^m$ with some positive integers $n,m$. 
We will see explicitly at 1-loop order that all $\beta$-functions of the other
couplings depend on $h_1$ indeed only through the combination $z h_1^2$. As a result it is meaningful to discuss
two different $z\to0$ limits: one where simply $z\to0$ is taken at finite $(g_1,g_2,g_3,g_4,h_1,h_2)$ 
and a second one where $z\to0$ in such a way that $zh_1^2$ is finite.

\section{RG flows with vector fields}
\label{rgflowswithvectorfields}

\begin{figure}
\begin{center}
\includegraphics[width=3.5cm]{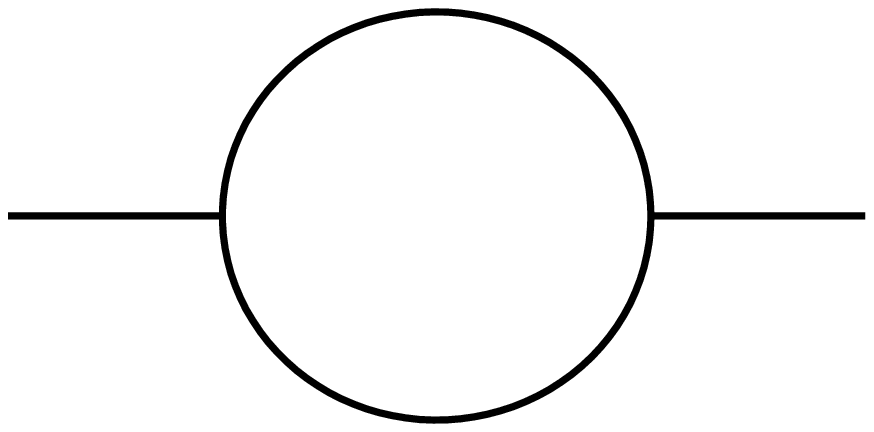} \includegraphics[width=3.5cm]{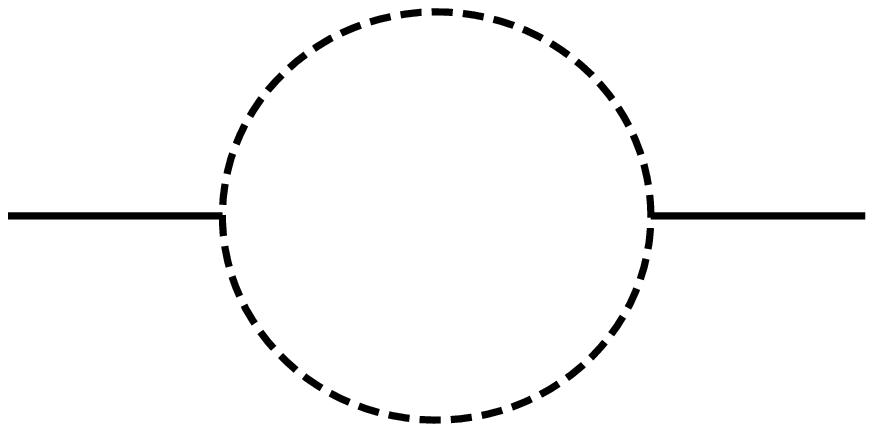} \\
\includegraphics[width=3.5cm]{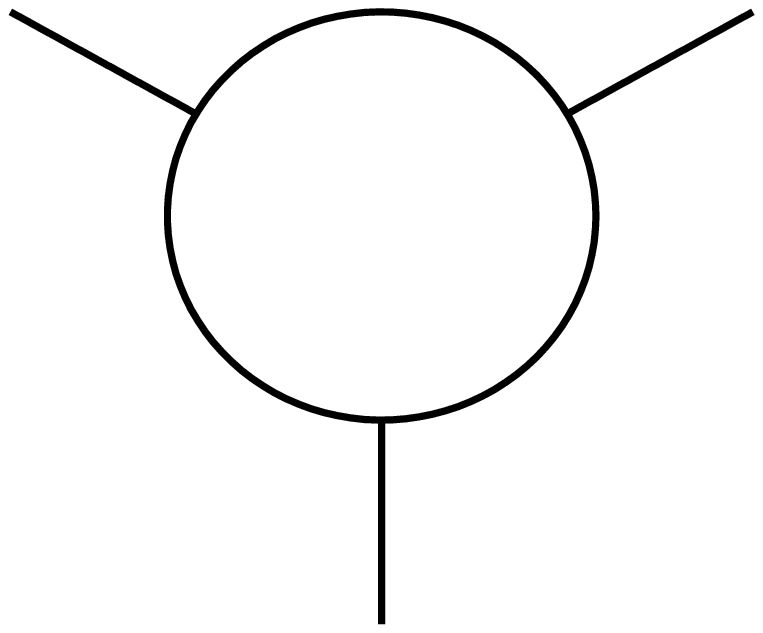} \includegraphics[width=3.5cm]{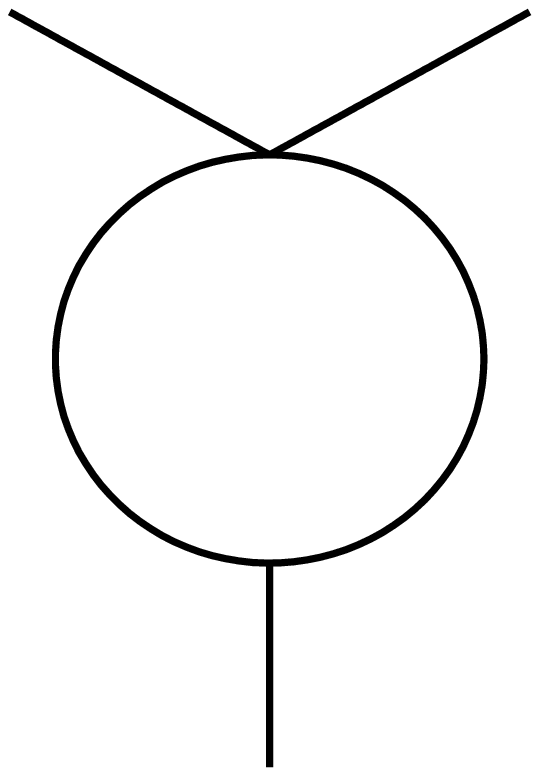} \includegraphics[width=3.5cm]{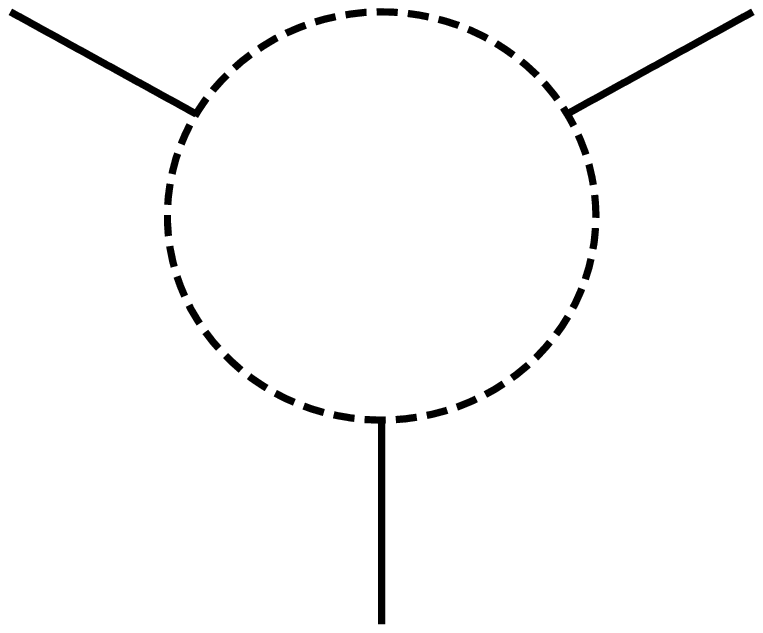} \\
\includegraphics[width=3.5cm]{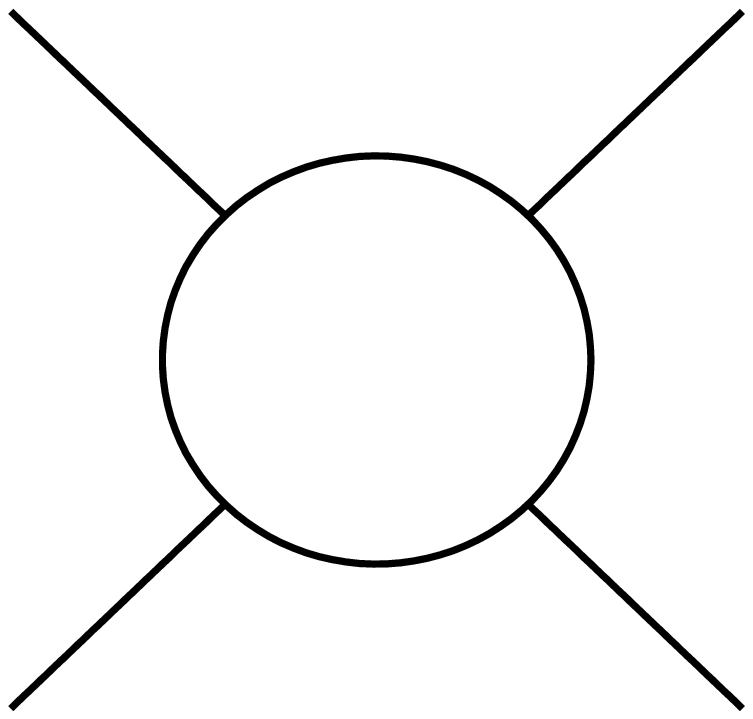} \includegraphics[width=3.5cm]{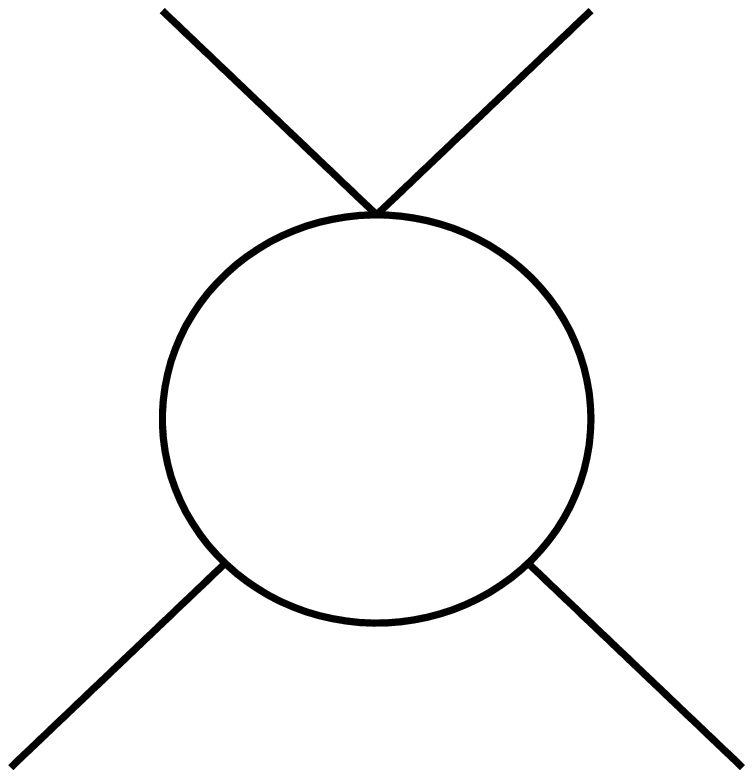} \includegraphics[width=3.5cm]{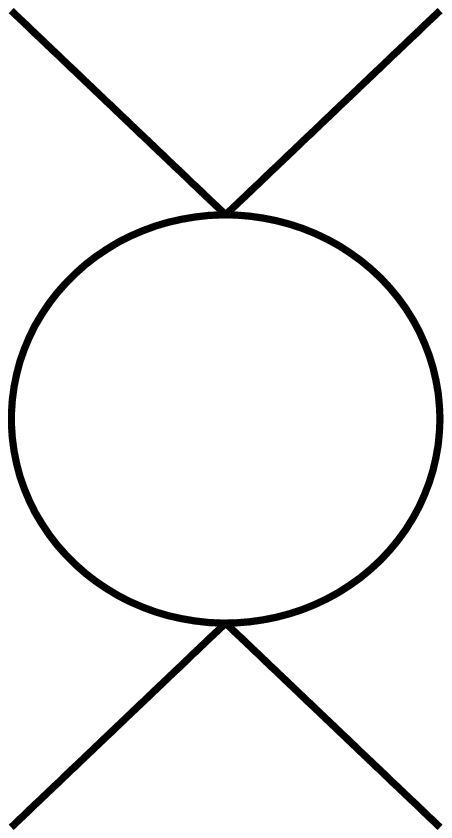}
    \includegraphics[width=3.5cm]{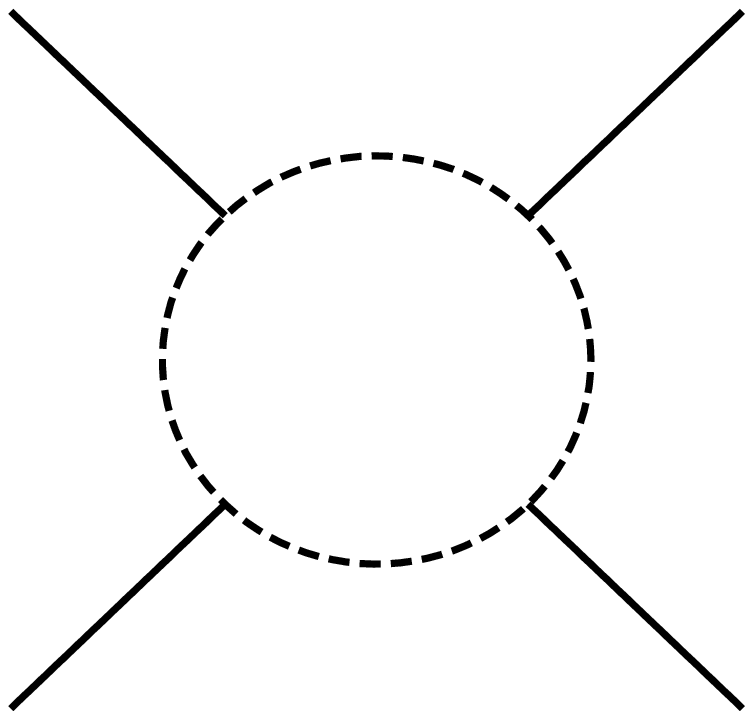} \\
\includegraphics[width=3.5cm]{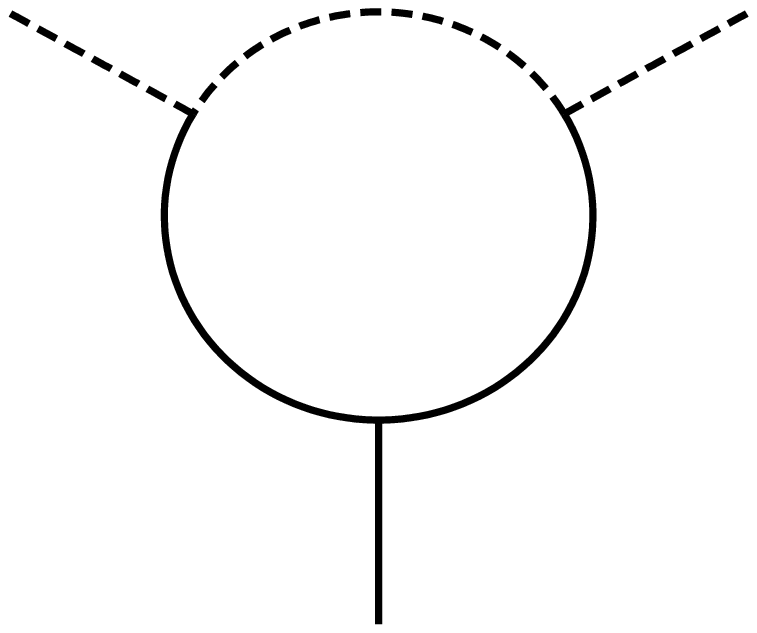} \includegraphics[width=3.5cm]{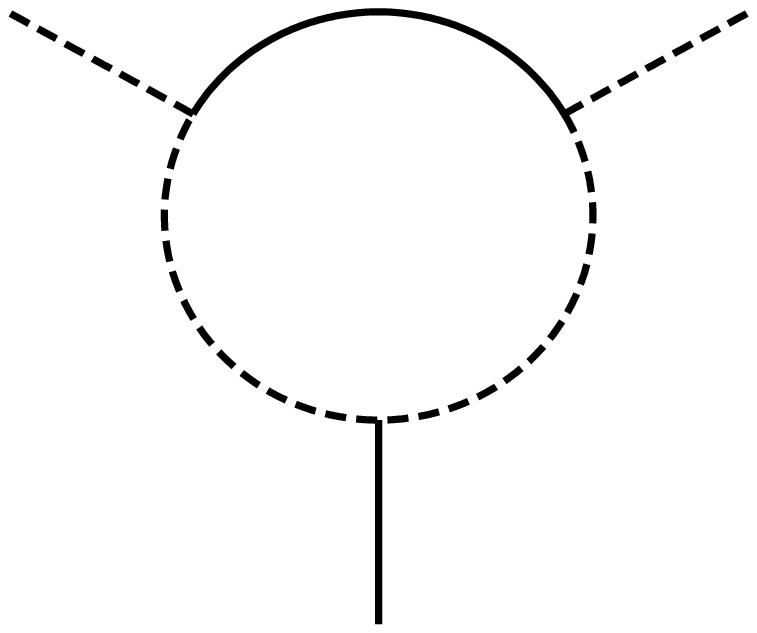}
\end{center}
\caption{Diagrams contributing at 1-loop order in dimensional regularization. 
Rows from top to bottom: propagator, renormalization of $z$;
3-vertex, renormalization of $(h_1,h_2)$; 4-vertex, renormalization of $(g_1,g_2,g_3,g_4)$;
ghost-vertex, renormalization of $(h_3,h_4)$. Solid and dashed lines represent vector and ghost fields, respectively.}
\label{diagrams}
\end{figure}

Now it is a straightforward exercise to compute the 1-loop $\beta$-functions in the space of our 7 couplings. 
The calculations were done with the help of FORM which turned out to be very useful in reducing the various group
theory factors \cite{Vermaseren:2000nd, Kuipers:2012rf, Ruijl:2017dtg}. In
this and subsequent sections all couplings will be assumed to be renormalized although the same notation will be used
as in the previous section for bare couplings. The
necessary diagrams are shown in figure \ref{diagrams}, including ghost fields, see section \ref{rgflows}. 
For simplicity let us introduce $(g_5,g_6) = ( h_1^2, h_2^2 )$
and work with the couplings $g_i$ with $i=1,\ldots,6$. 

We have in dimensional regularization \cite{tHooft:1972tcz, tHooft:1973mfk} and $\msbar$ scheme \cite{Bardeen:1978yd},
\bea
\label{beta1}
\mu \frac{dz}{d\mu} &=& \frac{1}{16\pi^2} \beta_z(z,g_5,g_6)  \\
\mu \frac{d g_i}{d \mu} &=& \frac{1}{16\pi^2} \beta_{g,i}(z,g_1,g_2,g_3,g_4,g_5,g_6)\;, \nn 
\eea
where $\beta_z$ is linear in $g_{5,6}$ and
the $\beta_{g,i}$ are quadratic in $g_j$. 
The specific form of $\beta_z$ and $\beta_{g,i}$ are fully given in
section \ref{rgflows} where an enlarged set of couplings and their $\beta$-functions are considered. One simply needs
to substitute $h_3 = h_4 = 0$ into the expressions (\ref{betas}) - (\ref{bbresult5}).
Note that $\beta_{g,5}$ is proportional to $g_5$ and $\beta_{g,6}$ 
is proportional to $g_6$ hence both couplings renormalize multiplicatively and so does $z$.
This completes our results for the 1-loop $\beta$-functions with vector fields only. 

Notice that if $g_{1,2,3}(\mu_0) = 0$ at some scale $\mu_0$, i.e. the renormalized Lagrangian 
corresponds to Yang-Mills theory, the RG-flow will generate $g_{1,2,3}(\mu) \neq 0$ at another scale 
$\mu \neq \mu_0$.

Just as the positivity of the classical potential is a non-trivial problem, it is similarly a non-trivial question
in general whether a classically positive potential stays positive under the 1-loop RG flow. The particular flows considered in
this paper will nonetheless have the property that if at some scale $\mu_0$ the potential is positive, then it will stay positive
for any scale $\mu > \mu_0$.

A simple cross-check may be performed at this stage on the $\beta$-functions. The
subspace \\ $(z,g_1,g_2,g_3,g_4,g_5,g_6) = (1,0,\lambda N/3, 0, 0, 0, 0, 0)$ corresponds to a $\lambda\phi^4$ theory with
$O(4(N^2-1))$ symmetry. Furthermore, the more general subspace $(z,g_1,g_2,g_3,g_4,g_5,g_6) = (1,g_1,g_2,g_3,g_4,0,0)$
corresponds to a general $\lambda\phi^4$ theory with $4(N^2-1)$ components 
and the known 1-loop $\beta$-functions can be recovered from (\ref{betas}) - (\ref{bbresult5}) in both cases.

\subsection{Asymptotically free RG flows}
\label{asymptoticallyfreetheories}

Clearly, we have a line of Gaussian fixed points given by $(z,g_1,g_2,g_3,g_4,g_5,g_6) = (z,0,0,0,0,0,0)$ for
arbitrary $z$.
We are primarily interested in asymptotically free quantum field theories with RG flows that towards
the UV approach a point on this line of Gaussian fixed points.

The coupling $z$ is different from the other couplings in the sense that the reliability of the perturbative expansion
requires $g_i \ll 1$ only, whereas $z$ is in principle arbitrary as long as it is finite. Fortunately, the 1-loop RG
flow does not produce $z\to\infty$ as a fixed point. This means that the delicate question of the $z\to\infty$ limit
does not arise at the 1-loop level. Henceforth we will assume $z$ is finite and only search for asymptotically free
theories with finite $z$.

If $g_5=g_6=0$ we are in
the subspace corresponding to $\lambda\phi^4$ type models and as such asymptotic freedom
is not possible. Hence we will assume that at least one of $(g_5,g_6)$ is non-zero.

It turns out there are two distinct possibilities for asymptotically free RG flows.
Following the discussion at the end of section \ref{lagrangianwithvectorfields} the dependence 
of the $\beta$-functions on $g_5 = h_1^2$ is only through the combination $zg_5$. In the first
type of RG flows we will set $g_5=0$ which also means $z g_5 = 0$ for any finite $z$.
Another possibility is $z = 0$ and $z g_5 \neq 0$ with finite $zg_5 / g_6$, 
in which case let us introduce $\tilde{g}_5 = zg_5$. 
We will search for asymptotically free RG flows in both setups. In any case $g_6 \neq 0$ is required.

It makes sense to introduce the ratios $r_i = g_i / g_6$ for $i=1,\ldots,4$ and $\tilde{r}_5 = z g_5 / g_6$, and
study the RG flow in the space $(z,r_1,r_2,r_3,r_4, g_5=0, g_6)$ and $(z=0,r_1,r_2,r_3,r_4,\tilde{r}_5,g_6)$
in the two aforementioned cases. So in the first case $g_5=0$ and in the latter $z=0$ at finite $\tilde{r}_5$.
We will be searching for non-trivial fixed points for $(z,r_1,r_2,r_3,r_4)$ and $(r_1,r_2,r_3,r_4,\tilde{r}_5)$,
respectively, and asymptotically free $g_6$. Once such flows are found, it is guaranteed that all original $g_i$  
couplings are asymptotically free as well. More specifically, we will find for $\mu\to\infty$,
\bea
\label{c6}
g_i(\mu) &\sim& 16\pi^2 \frac{C_i}{\log\frac{\mu}{\Lambda}}, \qquad i = 1,\ldots,4,6\;,
\eea
with some $\Lambda$ and similarly for $\tilde{g}_5(\mu)$ in the second case,
where the constant coefficients $C_{1,2,3,4}$ need to be checked on a case-by-case basis whether they correspond to a 
positive and stable potential ${\mathscr V}$ or not (\ref{lagrangian}). In any case $C_6 > 0$ 
is required since $g_6 = h_3^2 > 0$. Naturally, we have the constants $r_i = C_i / C_6$ for the ratios
and $\tilde{r}_5 = \tilde{C}_5 / C_6$ in the second setup.

Clearly, the UV fixed point is always at $g_i \to 0$ for $i=1,\ldots,4,6$ and $\tilde{g}_5 \to 0$ in the second setup. 
Nevertheless since we have more than one
coupling, the fixed point can be reached from various directions corresponding to various ratios $r_i$. In the space of
ratios a number of non-trivial fixed points exist, as we will see, and all these non-trivial fixed points correspond
to different quantum field theories. 

Naturally, if all couplings are asymptotically free, the 1-loop calculation is sufficient as usual. Higher loop
calculations are not necessary for establishing the existence of the well-defined perturbative quantum field
theories the non-trivial ratios define. 

Another natural way to introduce the ratios is by rescaling $A_\mu^a \to A_\mu^a / h_2$ and factoring out $g_6 = h_2^2$
in the Lagrangian,
\bea
{\mathscr L} = \frac{1}{g_6} \left( 
\frac{1}{2} \partial_\mu A_\nu^a \partial_\mu A_\nu^a - 
\frac{1}{2}\left(1-\frac{1}{z}\right) (\partial_\mu A_\mu^a )^2 + 
{\tilde{\mathscr O}}_2 + \frac{h_1}{h_2} {\tilde{\mathscr O}}_1 +
\sum_{i=1}^4 \frac{g_i}{g_6} {\mathscr O}_i 
\right) 
\label{nonpert}
\eea
which would be similar to the non-perturbative normalization of the field in Yang-Mills theory as opposed to the
perturbative normalization used in (\ref{lagrangian}). The ratios $h_1 / h_2 = \sqrt{r_5}$ and $g_i / g_6 = r_i$ 
are then directly the couplings in
(\ref{nonpert}) once $1 / g_6$ is factored out.

After these preliminary remarks let us find the fixed points for $(z,r_1,r_2,r_3,r_4)$ 
and $(r_1,r_2,r_3,r_4,\tilde{r}_5)$ in the two setups. Having the $\beta$-functions for all
couplings, it is in principle a straightforward exercise to search for ratios and $z$ such that they are fixed under the
RG flow. The solutions for $r_i$ (and $\tilde{r}_5$) are roots of polynomials with 
coefficients which are themselves polynomials in $N$. We
list the solutions in numerical form as well as $NC_6$ in 
table \ref{roots} and \ref{rootszg5}, respectively, for the two setups.

A number of remarks are in order.
There are two possible fixed points for $z$: either $z=0$ or $z=25/3$.
Interestingly, the $z=25/3$ fixed point is only present for $N < 6$. 
The fixed point $z=0$ exists however for all $N$ and persist in the large-$N$ limit as well;
see next section. The theory with $z=0$ is special in the sense that it corresponds to having only one kinetic term, the
first term in the Lagrangian (\ref{lagrangian}), but with the constraint $\partial_\mu A_\mu^a = 0$. Note that
gauge theory has not been mentioned at all up until this point since ghost fields have not been considered.
The constraint was obtained simply by studying the RG fixed points. In any case, the constraint reduces the number of
degrees of freedom from 4 to 3.

Once the fixed points are obtained it is worth investigating two issues. One, the stability of the potential ${\mathscr
V}$ and two, the stability of the fixed point in the RG sense. These two issues are of course completely independent.
For small $N$, i.e. $3 \leq N \leq 5$, there exist fixed points with both stable and unstable potentials, however for $N > 5$
all fixed points correspond to a stable ${\mathscr V}$. The RG stability of the results depends on the two setups we
have investigated, $g_5=0$ in the first case and $z=0$ with $\tilde{r}_5 \neq 0$ in the second case. In the first setup
at $N > 5$ there exist a unique fixed point such that the potential is stable and the fixed point is also stable in the
RG sense in the $z=0$ plane (where all fixed points lie). These are shown in bold in table \ref{roots}. In the second
setup all fixed points are RG unstable in at least one direction in the $z=0$ plane but all potentials are stable.

\begin{table}
\small
\begin{center}
\begin{tabular}{|c|c|c|c|c|c|c|c|c|}
\hline
$N$   & $z$   & $r_1$        & $r_2$        & $r_3$        & $r_4$        & $N C_6$      & ${\mathscr V}$ \\
\hline
\hline
    3 &     0 &     0.054652 &     0.122003 &     0.485317 &     0.970537 &     0.138656 &     stable \\ 
    3 &     0 &     0.064145 &     0.133021 &     0.665179 &     0.964086 &     0.137153 &     stable \\ 
    3 &     0 &    -0.647582 &    -0.580231 &     1.889786 &     1.204615 &     0.138656 &   unstable \\ 
    3 &     0 &    -0.562664 &    -0.493787 &     1.918797 &     1.173022 &     0.137153 &   unstable \\ 
    3 & 25/3  &     0.000334 &     0.079592 &    -0.251950 &     1.020083 &     0.148484 &   unstable \\ 
    3 & 25/3  &     0.010673 &     0.074642 &    -0.144563 &     1.004360 &     0.145542 &   unstable \\ 
    3 & 25/3  &    -0.108161 &    -0.028903 &    -0.034960 &     1.056248 &     0.148484 &   unstable \\ 
    3 & 25/3  &    -0.080316 &    -0.016348 &     0.037417 &     1.034690 &     0.145542 &   unstable \\ 
\hline
    4 &     0 &     0.044841 &     0.106784 &     0.351786 &     0.979028 &     0.140948 &     stable \\ 
    4 &     0 &     0.074162 &     0.083060 &     1.368389 &     0.960858 &     0.136196 &     stable \\ 
    4 & 25/3  &     0.004413 &     0.111209 &    -0.323177 &     1.013219 &     0.146900 &   unstable \\ 
    4 & 25/3  &     0.016297 &     0.243636 &    -0.344606 &     0.995511 &     0.145494 &   unstable \\ 
    4 & 25/3  &     0.017435 &     0.119096 &    -0.223217 &     0.997309 &     0.144605 &   unstable \\ 
    4 & 25/3  &     0.017931 &     0.235838 &    -0.327356 &     0.993784 &     0.145177 &   unstable \\ 
\hline
    5 &     0 &     0.042754 &     0.103223 &     0.327436 &     0.981138 &     0.141567 &     stable \\ 
    5 &     0 &     0.054311 &     1.073479 &     0.536511 &     0.957994 &     0.142046 &     stable \\ 
    5 &     0 &     0.067257 &    -0.066910 &     1.896637 &     0.967324 &     0.136857 &     stable \\ 
    5 &     0 &     0.069027 &     0.516675 &     1.600829 &     0.956705 &     0.138188 &     stable \\ 
    5 & 25/3  &     0.012566 &     0.149475 &    -0.375377 &     1.003344 &     0.145326 &   unstable \\ 
    5 & 25/3  &     0.021321 &     0.180212 &    -0.347564 &     0.993910 &     0.144298 &   unstable \\ 
\hline
\bf 6 & \bf 0 & \bf 0.041817 & \bf 0.101590 & \bf 0.316866 & \bf 0.982127 & \bf 0.141864 & \bf stable \\ 
    6 &     0 &     0.048648 &     1.137578 &     0.428569 &     0.966346 &     0.142530 &     stable \\ 
    6 &     0 &     0.059916 &    -0.214070 &     2.277709 &     0.972682 &     0.137748 &     stable \\ 
    6 &     0 &     0.062649 &     0.434621 &     2.043391 &     0.963808 &     0.138624 &     stable \\ 
\hline
\bf 7  & \bf 0 & \bf 0.041301 & \bf 0.100682 & \bf 0.311136 & \bf 0.982683 & \bf 0.142032 & \bf stable \\ 
    7  &     0 &     0.045944 &     1.161333 &     0.383774 &     0.971232 &     0.142626 &     stable \\ 
    7  &     0 &     0.054742 &    -0.321825 &     2.541816 &     0.976034 &     0.138570 &     stable \\ 
    7  &     0 &     0.057376 &     0.412019 &     2.341096 &     0.968497 &     0.139238 &     stable \\ 
\hline
\bf 10 & \bf 0 & \bf 0.040625 & \bf 0.099483 & \bf 0.303720 & \bf 0.983425 & \bf 0.142259 & \bf stable \\ 
    10 &     0 &     0.042691 &     1.184351 &     0.334451 &     0.977917 &     0.142606 &     stable \\ 
    10 &     0 &     0.047136 &    -0.495636 &     2.966144 &     0.980468 &     0.140207 &     stable \\ 
    10 &     0 &     0.048800 &     0.401667 &     2.839408 &     0.975942 &     0.140564 &     stable \\ 
\hline
\bf 50 & \bf 0 & \bf 0.040047 & \bf 0.098451 & \bf 0.297474 & \bf 0.984071 & \bf 0.142458 & \bf stable \\ 
    50 &     0 &     0.040124 &     1.198242 &     0.298567 &     0.983855 &     0.142474 &     stable \\ 
    50 &     0 &     0.040300 &    -0.680516 &     3.425269 &     0.983967 &     0.142360 &     stable \\ 
    50 &     0 &     0.040376 &     0.410710 &     3.418741 &     0.983752 &     0.142375 &     stable \\ 
\hline
\bf 100 & \bf 0 & \bf 0.040030 & \bf 0.098420 & \bf 0.297287 & \bf 0.984091 & \bf 0.142464 & \bf stable \\ 
    100 &     0 &     0.040049 &     1.198589 &     0.297559 &     0.984037 &     0.142468 &     stable \\ 
    100 &     0 &     0.040093 &    -0.686738 &     3.440904 &     0.984065 &     0.142439 &     stable \\ 
    100 &     0 &     0.040112 &     0.411281 &     3.439259 &     0.984011 &     0.142443 &     stable \\ 
\hline
$\mathbf\infty$ & \bf 0 & \bf 0.040024 & \bf 0.098409 & \bf 0.297224 & \bf 0.984097 & \bf 0.142466 & \bf stable \\
       $\infty$ &     0 &     0.040024 &     1.198704 &     0.297224 &     0.984097 &     0.142466 &     stable \\
       $\infty$ &     0 &     0.040024 &    -0.688818 &     3.446135 &     0.984097 &     0.142466 &     stable \\
       $\infty$ &     0 &     0.040024 &     0.411476 &     3.446135 &     0.984097 &     0.142466 &     stable \\
\hline
\end{tabular}
\end{center}
\caption{Non-trivial fixed points with $g_5 = 0$ for the ratios $r_i = g_i / g_6$,
and the coefficient $C_6$; see (\ref{c6}). The
last column indicate whether the potential ${\mathscr V}$ is stable or not.
For $N > 5$ there is a unique fixed point for which ${\mathscr V} \geq 0$ and is stable in the RG-sense
in the $z=0$ plane, these are shown in bold. 
}
\label{roots}
\end{table}

\begin{table}
\small
\begin{center}
\begin{tabular}{|c|c|c|c|c|c|c|}
\hline
$N$ & $r_1$    & $r_2$    & $r_3$    & $r_4$    & $\tilde{r}_5$ & $N C_6$ \\
\hline
\hline
3   & 1.346976 &  0.122003 & 0.485317 & 0.970537 & 1.292324 & 0.138656 \\
3   & 1.218191 &  0.133021 & 0.665179 & 0.964086 & 1.154046 & 0.137153 \\
3   & 0.644742 & -0.580231 & 1.889786 & 1.204615 & 1.292324 & 0.138656 \\
3   & 0.591382 & -0.493788 & 1.918797 & 1.173022 & 1.154046 & 0.137153 \\
\hline
4   & 1.124564 &  0.106784 & 0.351786 & 0.979029 & 1.079724 & 0.140948 \\
4   & 0.806856 &  0.083060 & 1.368389 & 0.960859 & 0.732694 & 0.136196 \\
\hline
5   & 1.032827 &  0.103223 & 0.327436 & 0.981138 & 0.990073 & 0.141567 \\
5   & 0.997387 &  1.073479 & 0.536511 & 0.957994 & 0.943076 & 0.142046 \\
5   & 0.750234 & -0.066910 & 1.896637 & 0.967324 & 0.682976 & 0.136857 \\
5   & 0.802230 &  0.516675 & 1.600828 & 0.956706 & 0.733204 & 0.138188 \\
\hline
6   & 0.988851 &  0.101590 & 0.316866 & 0.982127 & 0.947034 & 0.141864 \\
6   & 0.977346 &  1.137578 & 0.428569 & 0.966346 & 0.928698 & 0.142530 \\
6   & 0.757592 & -0.214070 & 2.277709 & 0.972682 & 0.697676 & 0.137748 \\
6   & 0.784051 &  0.434622 & 2.043391 & 0.963808 & 0.721403 & 0.138624 \\
\hline
7   & 0.964120 &  0.100682 & 0.311136 & 0.982683 & 0.922819 & 0.142032 \\
7   & 0.959302 &  1.161333 & 0.383774 & 0.971232 & 0.913358 & 0.142626 \\
7   & 0.777619 & -0.321825 & 2.541816 & 0.976034 & 0.722877 & 0.138570 \\
7   & 0.793631 &  0.412019 & 2.341096 & 0.968497 & 0.736254 & 0.139238 \\
\hline
10  & 0.931286 &  0.099483 & 0.303720 & 0.983425 & 0.890661 & 0.142259 \\
10  & 0.930723 &  1.184351 & 0.334451 & 0.977917 & 0.888032 & 0.142606 \\
10  & 0.827557 & -0.495636 & 2.966144 & 0.980468 & 0.780421 & 0.140207 \\
10  & 0.832435 &  0.401667 & 2.839408 & 0.975942 & 0.783635 & 0.140564 \\
\hline
50  & 0.902998 &  0.098451 & 0.297474 & 0.984071 & 0.862951 & 0.142458 \\
50  & 0.903023 &  1.198242 & 0.298567 & 0.983855 & 0.862899 & 0.142474 \\
50  & 0.898341 & -0.680516 & 3.425269 & 0.983967 & 0.858041 & 0.142360 \\
50  & 0.898375 &  0.410710 & 3.418741 & 0.983752 & 0.857999 & 0.142375 \\
\hline
100 & 0.902143 &  0.098420 & 0.297287 & 0.984091 & 0.862114 & 0.142464 \\
100 & 0.902150 &  1.198589 & 0.297559 & 0.984037 & 0.862101 & 0.142468 \\
100 & 0.900975 & -0.686738 & 3.440904 & 0.984065 & 0.860882 & 0.142439 \\
100 & 0.900982 &  0.411281 & 3.439259 & 0.984011 & 0.860870 & 0.142443 \\
\hline 
$\infty$ & 0.901859 &  0.098409 & 0.297224 & 0.984097 & 0.861835 & 0.142466 \\
$\infty$ & 0.901859 &  1.198704 & 0.297224 & 0.984097 & 0.861835 & 0.142466 \\
$\infty$ & 0.901859 & -0.688818 & 3.446135 & 0.984097 & 0.861835 & 0.142466 \\
$\infty$ & 0.901859 &  0.411476 & 3.446135 & 0.984097 & 0.861835 & 0.142466 \\
\hline
\end{tabular}
\end{center}
\caption{Non-trivial fixed points for the ratios $r_i = g_i / g_6$ and $\tilde{r}_5 = z g_5 / g_6$ with $z = 0$ 
but finite $\tilde{r}_5$. The coefficient $C_6$ from (\ref{c6}) is also shown. All of the corresponding
potentials are non-negative and hence stable but in the RG sense all fixed points are unstable in at 
least one direction.
}
\label{rootszg5}
\end{table}

To summarize the present section, we have found the surprising result that for any $N$ there are a finite number of 
fixed points for $z$ and the ratios of couplings leading to asymptotically free RG flows for all couplings. 
None of these are gauge theories. For $N>5$ there is
a unique fixed point which is both RG stable and gives rise to a stable potential ${\mathscr V}$. In
addition, all 8 fixed points for $N>5$ lie in the $z=0$ plane, which can be implemented by the constraint $\partial_\mu
A_\mu^a = 0$. The features found for $N > 5$ persist in the large-$N$ limit which will be the subject of the next
section.

\subsection{Large-$N$ limit}

Tables \ref{roots} and \ref{rootszg5} show explicitly that a smooth large-$N$ limit exists once the $N\to\infty$ limit is
performed at finite $Ng_i$ for $i=1,\ldots,6$. In terms of the original $h_{1,2}$ couplings in the Lagrangian this of
course means constant $\sqrt{N} h_{1,2}$. Searching for fixed points for $z$ and the ratios
follows the same path as the finite $N$ calculation only the equations are somewhat simpler. The fixed points are still
roots of high order polynomials and the only fixed point for $z$ is $z=0$ leading to $\partial_\mu A_\mu^a = 0$. There
are 8 fixed points, 4 for which $g_5 = 0$ and another 4 with finite $z g_5 / g_6 = \tilde{r}_5$ but still $z=0$.
The $N C_6$ coefficient determining the RG-flow of $N g_6$ is the same for all 8 fixed points. Furthermore,
all 8 fixed points correspond to a
positive semi-definite potential ${\mathscr V} \geq 0$. And there is a unique fixed point which is fully RG stable, 
this is shown in bold in table \ref{roots}. 
All of these features were present for finite $N > 5$ the only new aspect of the strict large-$N$ limit is the
degeneracy in some of the couplings among the 8 solutions. 

In any case the resulting 8 theories are well-defined local, interacting, asymptotically free, 
Euclidean quantum field theories in four dimensions.

\section{RG flows with vector fields and ghosts}
\label{rgflows}

In this section we enlarge the space of couplings to include ghost fields. The
motivation to do so is to establish how perturbative gauge theories fit into the general phase space of global $SU(N)$
invariant theories. If we were to start with Yang-Mills theory, gauge fixing would of course be required for any
perturbative calculation which in turn would
introduce gauge fixing terms in the Lagrangian as well as ghosts\footnote{We limit ourselves to the Lorentz gauge.}. 
Hence this perturbative gauge theory situation would be included in our most general setup if we included the most
general Lagrangian with ghosts since the kinetic terms already account for the Lorentz gauge condition. 

There are two possible ghost Lagrangians with global $SU(N)$ symmetry,
\bea
{\mathscr L}_{ghost} = \partial_\mu {\bar c}^a \left( \delta_{ac} \partial_\mu  
+ h_3 d_{abc} A_\mu^b + h_4 f_{abc} A_\mu^b \right) c^c 
\eea
assuming canonical normalization for the ghosts fields $c^a$, leading to two new couplings, $h_3$ and $h_4$.
Hence over-all we have a 9 dimensional space of couplings $(z,g_1,g_2,g_3,g_4,h_1,h_2,h_3,h_4)$. Sometimes we use the
notation $(g_5,g_6,g_7,g_8) = ( h_1^2, h_2^2, h_3^2, h_4^2 )$ as well. 

The diagrams involving ghosts are shown also in figure \ref{diagrams} and we are led to the following
$\beta$-functions,
\bea
\label{betas}
\mu \frac{dz}{d\mu} &=& \frac{1}{16\pi^2} \beta_z(z,g_5,g_6,g_7,g_8) \nn \\
\mu \frac{d g_i}{d \mu} &=& \frac{1}{16\pi^2} \beta_{g,i}(z,g_1,g_2,g_3,g_4,g_5,g_6,g_7,g_8) \qquad i=1,2,3,4 \\
\mu \frac{d h_i}{d \mu} &=& \frac{1}{16\pi^2} \beta_{h,i}(z,g_1,g_2,g_3,g_4,h_1,h_2,h_3,h_4) \qquad i=1,2,3,4\;. \nn
\eea
The $\beta_{g,i}$ are quadratic in $g_1,\ldots,g_8$ and are parametrized as
\bea
\label{bbresult2}
\beta_{g,i} &=& \frac{N}{24} B_i^{jk} g_j g_k \nn \\
B_i &=& B^{(0)}_i + \frac{1}{N^2} B^{(1)}_i\;,
\eea
where the $B^{(0)}_i$ and $B^{(1)}_i$ matrices are $8\times 8$, depend on $z$ and are given in the appendix. Note that
no approximation was made at this point, the separation into $O(1)$ and $O(1/N^2)$ terms is exact.
The result for $\beta_z$ is given by,
\bea
\label{bbresult1}
\beta_z &=&  - \frac{z}{6} N \left( 
g_5 \frac{N^2-4}{N^2} z ( 15z^2-19z+18 ) + g_6 (3z-25) 
- g_7 \frac{N^2-4}{N^2} (3z-1) + g_8 (3z-1)
\right)
\eea
and finally the $\beta$-functions of the cubic and ghosts couplings are
\bea
\beta_{h,1} &=& \frac{N}{12} \left( 
h_3 \left( 3 \frac{12-N^2}{N^2} g_7 + 9 g_8  \right) + 
h_1 \, g_j \, b_1^j 
\right) \nn \\
\beta_{h,2} &=& \frac{N}{12} \left( 
h_4 \left( 3 \frac{4-N^2}{N^2} g_7 + g_8  \right) + 
h_2 \, g_j \, b_2^j 
\right) \nn
\eea
\bea
\label{bbresult5}
b_1 = \lmatrix{c}
3 \left(8 z^{2} - 5 z + 15\right) - \frac{36 \left(4 z^{2} - z + 7\right)}{N^{2}} \\
\frac{6 \left(2 z^{2} + z + 3\right)}{N^{2}} \\
\frac{3 \left(8 z^{2} - 5 z + 15\right)}{N^{2}} \\
3 \left(4 z^{2} - 7 z + 9\right) \\
- 3 z \left(5 z + 3\right) + \frac{12 z \left(13 z + 9\right)}{N^{2}} \\
- 3 \left(4 z^{2} - 7 z + 37\right) \\
3 - \frac{12}{N^{2}} \\
-3 
\rmatrix
\qquad
b_2 = \lmatrix{c}
3(z+5)\left(1-\frac{4}{N^2}\right) \\
6\frac{z+5}{N^2} \\
-3\frac{z+5}{N^2} \\
9(z+5) \\
-3z(z+5) \left( 1  - \frac { 4 }{N^2} \right) \\
-3(3z+29) \\
3-\frac{12}{N^2} \\
-3 \rmatrix 
\eea
\bea
\beta_{h,3} &=& \frac{N}{12} \left(
9 h_1 h_4^2 z
-h_1^2h_3 z^2 \frac {N^2-4}{N^2}
-9h_1h_3^2 z \frac{ N^2-12}{N^2} + \right. \nn \\
& & \left. + h_2^2h_3 (6z-25)
-18 h_2 h_3 h_4 z
- h_3^3 \left( 3z-19 + 4 \frac{ 3z+19 }{N^2} \right)
+ h_3 h_4^2 ( 9z-19 ) 
\right) \nn \\
\beta_{h,4} &=& \frac{N}{12} \left(
9 h_2 h_3^2 z \frac{N^2-4}{N^2}
- h_1^2 h_4 z^2 \frac { N^2-4 }{N^2}
-18 h_1 h_3 h_4 z \frac { N^2-4  }{N^2} + \right. \nn \\
& & \left. + h_2^2 h_4 ( 6z-25 ) 
- 9 h_2 h_4^2 z 
- h_3^2 h_4 ( 9z-19 ) \frac{ N^2-4 }{N^2} 
+ h_4^3 ( 3z-19 ) \nn
\right)\;.
\eea
This concludes our calculation of the 1-loop $\beta$-functions of all 9 couplings.

\subsection{Asymptotically free RG flows and emergent gauge theory}

Let us see how perturbative $SU(N)$ gauge theory in Lorentz gauge fits into the general construction.
This setup corresponds to $(z,g_1,g_2,g_3,g_4,h_1,h_2,h_3,h_4) = (z,0,0,0,g^2,0,g,0,g)$ where $z$ is the gauge fixing
parameter and $g$ is the usual gauge coupling. Substituting into the above $\beta$-functions results in,
\bea
\label{qcd}
\begin{array}{llll}
\beta_{g,1} = 0\;,\quad & \beta_{g,2} = 0\;,\quad                 & \beta_{g,3} = 0\;,\quad & \beta_{g,4} = -\frac{22}{3}Ng^4 \\
& & & \\
\beta_{h,1} = 0\;,\quad & \beta_{h,2} = -\frac{11}{3}Ng^3\;,\quad & \beta_{h,3} = 0\;,\quad & \beta_{h,4} = -\frac{11}{3}Ng^3\;,
\end{array}
\eea
hence we correctly reproduce the known $\beta$-function from the 3-gluon, 4-gluon and ghost-gluon vertices and the 
remaining $\beta$-functions are zero, consistent with the choices $g_1=g_2=g_3=h_1=h_3=0$. The $\beta$-function
for the gauge fixing parameter $z$ (the notation $\xi = 1-z$ is often used) is easily obtained as well from (\ref{bbresult1}),
\bea
\beta_z = z \frac{13-3z}{3} g^2 N\;,
\eea
which reproduces the known result \cite{Gross:1973ju}. Note that from the point of the view of the general
9-dimensional coupling space, the fact that the $\beta$-functions (\ref{qcd}) do not depend on $z$ is a result
of non-trivial cancellations.
In general the $\beta$-functions do depend on $z$ and the lack of this dependence is a peculiarity of the
gauge theory special case.

Clearly, the Yang-Mills corner of the parameter space can be reached by tuning a finite number of couplings.
For generic values of the couplings we only have global $SU(N)$ invariance. By tuning a finite number of them
we end up with an infinite dimensional local $SU(N)$ invariance. Viewed this way, gauge symmetry is an
emergent phenomenon specific to a particular RG flow. Generally, particular corners of the coupling space frequently
have larger symmetries than a generic point, but in our situation the emergent symmetry is infinite dimensional
although we are dealing with a finite number of couplings. Another emergent feature of the Yang-Mills corner
of the coupling space is unitarity. The theories corresponding to RG flows different from gauge theory (see below)
are Euclidean and do not admit a unitary Minkowskian version due to the lack of gauge invariance taking 
care of the negative norm states. Only the RG flow corresponding to Yang-Mills can be continued to a unitary
Minkowskian quantum field theory.

\begin{table}
\small
\begin{center}
\begin{tabular}{|c|c|c|c|c|c|c|c|c|c|}
\hline
$z$ & $r_1$ & $r_2$ & $r_3$ & $r_4$ & $R_5$ & $R_7$ & $R_8$ & $N C_6$ & ${\mathscr V}$ \\
\hline
\hline
0   &   0   &   0   &   0   &   1   &   0   &   0   &   1   & 3/22 & stable \\
\hline
0   &   0   &   -1  &   4   &   1   &   0   &   0   &   1   & 3/22 & stable \\
\hline
0   &   0   &   4/9 &   4   &   1   &   0   &   0   &   1   & 3/22 & stable \\
\hline
0   &   0   &  13/9 &   0   &   1   &   0   &   0   &   1   & 3/22 & stable \\
\hline
13/3&   0   &   0   &   0   &   1   &   0   &   0   &   1   & 3/22 & stable \\
\hline
13/3&   0   & 39/98 &   0   &   1   &   0   &   0   &   1   & 3/22 & stable \\
\hline
13/3&   0   & 27/194  &-54/97&  1   &   0   &   0   &   1   & 3/22 & unstable \\
\hline
13/3&   0   &2553/4753&-54/97&  1   &   0   &   0   &   1   & 3/22 & unstable \\
\hline
3   &  1/9  &  7/9  & -2/3  &   1   &$1/\sqrt{27}$&$1/\sqrt{3}$&   1   & 3/22 & stable \\
\hline                                                         
3   &  1/9  &   0   &   0   &   1   &$1/\sqrt{27}$&$1/\sqrt{3}$&   1   & 3/22 & stable \\
\hline                                                         
3   &  1/9  &  1/6  & -2/3  &   1   &$1/\sqrt{27}$&$1/\sqrt{3}$&   1   & 3/22 & stable \\
\hline                                                         
3   &  1/9  & 11/18 &   0   &   1   &$1/\sqrt{27}$&$1/\sqrt{3}$&   1   & 3/22 & stable \\
\hline
... &       &       &       &       &             &            &       &      &    \\
\hline
\end{tabular}
\end{center}
\caption{Non-trivial fixed points in the large-$N$ limit for $z$, the ratios $r_{1,2,3,4} = g_{1,2,3,4} / h_2^2, R_5=h_1/h_2, 
R_7 = h_3/h_2, R_8 = h_4/h_2$ and the coefficient $N C_6$ determining the UV running of $g_6 = h_2^2$; see (\ref{c6}).
The first and fifth lines correspond to gauge theory, they only differ by the value of $z$ which is clearly 
    irrelevant in this case. The large-$N$ limit of the finite $N$ solution found in (\ref{analytic})
    is the $10^{th}$ line above. The $\ldots$ represent further 24 solutions found numerically.} 
\label{largen}
\end{table}

Beyond the gauge theory special case there are however other asymptotically free RG flows.
Following the calculation in section \ref{asymptoticallyfreetheories} we determine these next.
We will be searching for fixed points for $z$ and the ratios
\bea
r_{1,2,3,4} = \frac{g_{1,2,3,4}}{h_2^2}\,, \quad  R_5 = \frac{h_1}{h_2}\,, \quad  R_7 = \frac{h_3}{h_2}\,, \quad
R_8 = \frac{h_4}{h_2} \nn
\eea
and asymptotically free $g_6(\mu) = h_2^2(\mu)$. Once such fixed points are found, all couplings are
asymptotically free towards the UV. Yang-Mills theory corresponds to $r_{1,2,3} = R_{5,7} = 0$, $R_{4,8} = 1$ and
$N C_6 = 3/22$, the familiar coefficient of the 1-loop $\beta$-function, see (\ref{c6}).

Finding the fixed points for $z$ and the 7 ratios $r_i, R_{5,7,8}$ amounts to again solving a system of polynomial
equations. 
A large number of fixed points are found for any $N$, most of them numerically. We have found one example (beyond the
gauge theory case) which is exact at finite $N$ as well and is listed below together with $NC_6$ determining the running
of $g_6$,
\bea
\label{analytic}
\begin{array}{ccccc}
    z = 3\;,\; & r_1 = \frac{N^2}{9(N^2-4)}\;,\; & r_2 = 0\;, & r_3 = 0\;,\; & r_4 = 1 \\ \\
    & R_5 = \frac{N}{\sqrt{27(N^2-4)}}\;,\; & R_7 = \frac{N}{\sqrt{3(N^2-4)}}\;,\; & R_8 = 1\;,\; & NC_6 =
    \frac{3}{22}\;.
\end{array}
\eea
This example only differs from the gauge theory case by $r_1, R_{5,7}$ and $z$, hence can be considered
a perturbation by 3 marginal operators.
Instead of listing all other fixed points for the ratios which are found numerically, 
we focus on the large-$N$ limit which illustrates the phase structure and is somewhat
simpler. 

\subsection{Large $N$-limit}

It turns out 12 of the fixed points in the large $N$-limit can even be found analytically, 
these are listed in table \ref{largen}. 
In addition there are 24 more fixed points, found numerically, but since they are not very illuminating they are 
not listed explicitly.

Notice that the 12 fixed points correspond to the same RG flow of $g_6(\mu)$ given by the same $NC_6 = 3/22$
coefficient (\ref{c6}) as in Yang-Mills theory. Two of the 12 fixed points 
are precisely Yang-Mills theory, namely the first and the fifth row of table \ref{largen}. The two are distinguished
only by the fixed point value of $z$, in one case $z=0$ in the other $z=13/3$. The former is unstable in the RG sense
in the $z$-direction, the latter is stable.
Eight of the other 10 fixed points correspond to a non-negative and hence
stable potential ${\mathscr V}$ and as such lead to completely well-defined asymptotically free 
marginal perturbations of Euclidean gauge theory.

\subsection{Perturbation by double trace operators}

\begin{figure}
\begin{center}
\includegraphics[height=5.2cm]{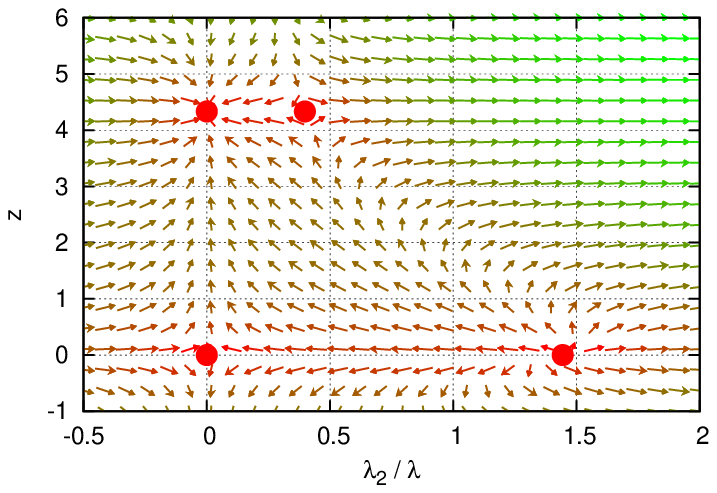} \includegraphics[height=5.2cm]{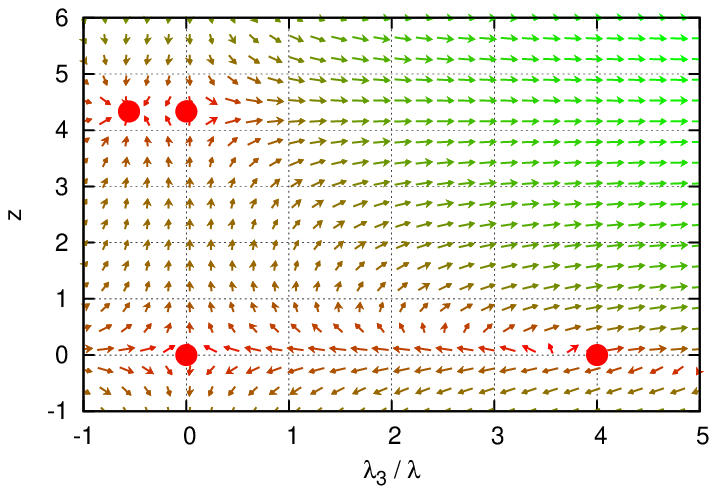} 
\end{center}
\caption{The RG flows corresponding to two gauge symmetry breaking perturbations of 
Yang-Mills theory in the large-$N$ limit. Left: the $(\lambda_2/\lambda,z)$ plane at 
$\lambda_3 = 0$, right: $(\lambda_3/\lambda,z)$ plane
at $\lambda_2 = 0$. The fixed points are shown with solid red dots. The vertical axis
at $\lambda_2/\lambda = 0$ and $\lambda_3/\lambda=0$ on the left and right, respectively, correspond to Yang-Mills
theory with $z=0$ and $z=13/3$,
the fixed points away from these vertical lines correspond to non-trivial RG flows.}
\label{def}
\end{figure}

The first 8 of the fixed points only differ in $(z,r_2,r_3)$ and are all in the subspace $r_1 = R_{5,7} = 0, r_4 = R_8 = 1$.
Since Yang-Mills theory also corresponds to $r_1 = R_{5,7} = 0, r_4 = R_8 = 1$ 
it is natural to consider the space of 4 couplings
$(z,g,g_2,g_3)$ defining a 2-parameter marginal perturbation, where $g$ is the Yang-Mills coupling. 
Gauge theory only involves $(z,g)$ and the remaining
$(g_2,g_3)$ correspond to the double trace operators ${\mathscr O}_2$ and ${\mathscr O}_3$ (\ref{lagrangian}).
For convenience let us introduce 
\bea
\kappa_2 = g_2 + \frac{1}{4} g_3\;, \qquad \qquad   \kappa_3 = g_3
\eea
Then we are led to study the following 2-parameter marginal perturbation of Yang-Mills theory,
\bea
{\mathscr L} &=& - \frac{1}{4} \tr F_{\mu\nu} F_{\mu\nu} + {\mathscr L}_{gf} + {\mathscr L}_{ghost} + 
\kappa_2 {\mathscr O}_2  + \kappa_3  \left( {\mathscr O}_3 - \frac{1}{4} {\mathscr O}_2 \right) \nn \\
{\mathscr L}_{gf} &=& \frac{1}{2z} ( \partial_\mu A_\mu^a )^2 \nn \\ 
{\mathscr L}_{ghost} &=& \partial_\mu {\bar c}^a \left( \delta_{ac} \partial_\mu  
+ g f_{abc} A_\mu^b \right) c^c \\
{\mathscr O}_2 &=& \frac{1}{8N}( A_\mu^a A_\mu^a )^2 = \frac{1}{2N} \left( \tr( A_\mu A_\mu ) \right)^2  \geq 0 \nn \\
{\mathscr O}_3 &=& \frac{1}{8N} A_\mu^a A_\mu^b A_\nu^a A_\nu^b = \frac{1}{2N} \tr(A_\mu A_\nu) \tr(A_\mu A_\nu) \geq 0
\nn \\
{\mathscr O}_3 - \frac{1}{4} {\mathscr O}_2 &\geq& 0 \nn
\eea
in the large-$N$ limit. Introducing $\lambda = g^2 N$, $\lambda_2 = \kappa_2 N$ and $\lambda_3 = \kappa_3 N$ for the 't Hooft
couplings we have explicitly, for the 1-loop $\beta$-functions,
\bea
\beta_{\lambda_2} &=& \lambda_2 \left(  \lambda_2 \left(z^2+3\right)   - \lambda \left(z+\frac{35}{3}\right) \right) \nn \\
\beta_{\lambda_3} &=& \frac{1}{12} \lambda_3 \left( \lambda_3 ( z^2+4z+7 )  - \lambda (116-12z) \right) \nn \\
\beta_{\lambda} &=& - \frac{22}{3} \lambda^2 \\
\beta_z &=& z \frac{13-3z}{3} \lambda \nn \;.
\eea
Clearly, the marginal perturbation by $(\lambda_2,\lambda_3)$ leaves the RG flow of the gauge coupling and $z$
the same as without the perturbation, at least to 1-loop order. Note that $\lambda_2$ and $\lambda_3$ can 
independently be switched on or off, the flow of one does not involve the other. Furthermore, both renormalize
multiplicatively. It is a simple exercise to verify the first
8 fixed points for the ratios $r_{2,3} = g_{2,3} / g^2$ and $z$ listed in table \ref{largen}. 

For illustration the RG flow of $(\lambda_2/\lambda,z)$ at $\lambda_3 = 0$ and of $(\lambda_3/\lambda,z)$ at $\lambda_2 = 0$ are
shown in figure \ref{def}. The fixed points are shown by solid red dots. Yang-Mills corresponds to $\lambda_2 =
\lambda_3 = 0$ and either $z=0$ or $z=13/3$. The latter is stable, the former is unstable in the $z$-direction.

Notice that Yang-Mills (with $z=13/3$) is unstable in the $\lambda_3$ direction, 
there is a non-trivial stable fixed point $\lambda_3 / \lambda = -54/97$ and $\lambda_2 = 0$, $z=13/3$. This point
is at the end of the RG trajectory starting from Yang-Mills with an arbitrarily small (negative) 
$\lambda_3 / \lambda$ perturbation. The potential ${\mathscr V}$ is however unstable for negative $\lambda_3$. 

To summarize, out of the 6 non-trivial fixed points listed in table \ref{largen} 
which are not Yang-Mills theory and belong to the 2-parameter perturbation considered in this section, 
4 give rise to a stable potential. These are only partially stable in the RG sense.

\section{Conclusion and outlook}
\label{conclusion}

In this paper the most general local, classically scale invariant, perturbatively renormalizable,
globally $SU(N)$ invariant 4 dimensional Euclidean quantum field theory of vector fields was investigated. 

There are 2 main results. First, if $N>5$ we have found that there is a unique RG flow towards the UV which is
asymptotically free, corresponds to a non-negative stable potential and is also RG stable. This particular
RG flow defines a perfectly well-defined Euclidean quantum field theory in four dimensions without
gauge invariance. There are additional RG flows which are asymptotically free, correspond to a non-negative stable
potential but are only partially stable in the RG sense. These can be realized by some amount of fine tuning. 

The second set of results concerns the coupling of vector fields and ghosts. The setup naturally includes perturbative
Yang-Mills theory and one can explicitly see how infinite dimensional gauge invariance emerges on a particular asymptotically
free RG flow but not on others. Concretely, a general point in the space of couplings has a finite dimensional group of
symmetries but by tuning a finite number of couplings, an infinite dimensional invariance can be recovered. In this
sense gauge symmetry is an emergent phenomenon, specific to a particular RG flow. On this RG flow unitarity is also
emergent, since, as is well-known, gauge invariance is required for a unitary 
and Lorentz invariant theory of vector fields
in Minkowskian signature. 
Some of the other RG flows can naturally be identified as marginal perturbations of
Yang-Mills theory and in the large-$N$ limit particularly simple expression were obtained.

It turned out both with and without ghosts that the $\partial_\mu A_\mu^a = 0$ constraint 
naturally arises from the study
of the general RG flows. Namely, if asymptotic freedom is required with the same logarithmic running, $\sim 1 /
\log(\mu/\Lambda)$ for all couplings, then ratios of couplings are constant 
and define non-trivial fixed points in the UV. Requiring a fixed point
for the quadratic $z$ coupling as well leads to $z=0$ for $N>5$
which in turn is equivalent to the aforementioned constraint.
Hence the reduction of degrees of freedom from 4 to 3 is a dynamical phenomenon. Further reduction to 2 degrees of
freedom only occurs at the particular RG flow corresponding to gauge theory.

Classically scale invariant Lagrangians were considered only, i.e. dimensionless couplings only. There is
however one coupling, the mass term for the vector field,
\bea
{\mathscr L}_m = \frac{m^2}{2} A_\mu^a A_\mu^a
\eea
which is still renormalizable and could be included as well. It breaks gauge invariance, 
similarly to the general dimensionless couplings
considered already. A consistent way to exclude the above mass term is in dimensional regularization, as done in this
paper, but with other regulators it should be included in general. Turn it around, if one starts from gauge theory and
gauge symmetry is broken by some mechanism, a mass term is usually generated. Once this is the case there is no reason
to exclude the other (dimensionless) gauge invariance breaking couplings considered in this paper.

All computations were done on the 1-loop level and it would be very interesting to extend them to higher loops. Certain
aspects are not expected to change, namely the existence of asymptotically free RG flows towards the UV,
since these can be established at $g_i \ll 1$ and $\mu\to\infty$ where higher loop corrections can be made arbitrarily
small.

It would be interesting to repeat the calculation of the $\beta$-functions in a setup where Lorentz symmetry is broken
by a preferred direction $n_\mu$. Again a finite number of marginal couplings exist with global $SU(N)$ invariance. 
Yang-Mills with gauge choices without ghosts, such as the axial gauge, would be naturally included in this setup
as particular corners of the coupling space. At the fixed point corresponding to Yang-Mills, Lorentz symmetry
would be emergent and it would be interesting to study the behavior at other non-trivial fixed points, if any
existed.

One of the motivation for the present work was the following analogy with quantum gravity: in any acceptable quantum
theory of gravity one expects diffeomorphism invariance, which is an infinite dimensional invariance. There are
formulations which start from a regularized setup where diffeomorphisms are 
broken. It is hoped that in some limit (e.g. continuum
limit if a lattice regulator is used \cite{Catterall:2009nz}) it is recovered. 
It is entirely possible that these approaches are sufficiently
general so that they include several continuum theories as distinct fixed points and only some of them correspond to the
desired infinite dimensional invariance. In our work this was shown to be possible, although of course not with spin-2
fields, without fine tuning an infinite number of couplings.

\appendix

\section{1-loop $\beta$-functions}

The $\beta_{g,i}$ functions are given by the $B^{(0)}$ and $B^{(1)}$ matrices (\ref{bbresult2}). These are 
listed below.

\begin{landscape}
\small
\bea
\label{bbresult3}
B^{(0)}_1 &=& \lmatrix{cccccccc}
3 \left(13 z^{2} + 2 z + 41\right) & 0 & 0 & 3 \left(7 z^{2} - 2 z + 19\right) & - z \left(25 z + 33\right) & - 21 z^{2} + 3 z - 184 & 4 & -4  \\
0 & 0 & 0 & 0 & 0 & 0 & 0 & 0  \\
0 & 0 & 0 & 0 & 0 & 0 & 0 & 0  \\
3 \left(7 z^{2} - 2 z + 19\right) & 0 & 0 & 3 \left(5 z^{2} + 2 z + 17\right) & - 3 z \left(5 z + 1\right) & - 15 \left(z^{2} + z + 4\right) & 0 & 0  \\
- z \left(25 z + 33\right) & 0 & 0 & - 3 z \left(5 z + 1\right) & 15 z^{2} & 3 z \left(5 z + 4\right) & 0 & 0  \\
- 21 z^{2} + 3 z - 184 & 0 & 0 & - 15 \left(z^{2} + z + 4\right) & 3 z \left(5 z + 4\right) & 3 \left(5 z^{2} + 8 z + 24\right) & 0 & 0  \\
4 & 0 & 0 & 0 & 0 & 0 & -3 & 9  \\
-4 & 0 & 0 & 0 & 0 & 0 & 9 & -3  \\
\rmatrix \nn \\ \nn \\
B^{(0)}_2 &=& \lmatrix{cccccccc}
12 \left(z^{2} + 2 z + 5\right) & 12 \left(z^{2} + 3\right) & 2 \left(z - 1\right)^{2} & 4 \left(5 z^{2} + 2 z + 17\right) & - 12 z \left(z + 5\right) & - 20 \left(z^{2} + z + 4\right) & 0 & 0  \\
12 \left(z^{2} + 3\right) & 24 \left(z^{2} + 3\right) & 6 \left(z^{2} + 3\right) & 36 \left(z^{2} + 3\right) & - 4 z \left(10 z + 9\right) & - 4 \left(9 z^{2} + 3 z + 61\right) & 4 & -4  \\
2 \left(z - 1\right)^{2} & 6 \left(z^{2} + 3\right) & \left(z - 1\right)^{2} & 2 \left(5 z^{2} + 2 z + 17\right) & - 2 z \left(5 z + 1\right) & - 10 \left(z^{2} + z + 4\right) & 0 & 0  \\
4 \left(5 z^{2} + 2 z + 17\right) & 36 \left(z^{2} + 3\right) & 2 \left(5 z^{2} + 2 z + 17\right) & 12 \left(5 z^{2} + 2 z + 17\right) & - 4 z \left(13 z + 17\right) & - 60 \left(z^{2} + z + 4\right) & 0 & 0  \\
- 12 z \left(z + 5\right) & - 4 z \left(10 z + 9\right) & - 2 z \left(5 z + 1\right) & - 4 z \left(13 z + 17\right) & 60 z^{2} & 4 z \left(13 z + 20\right) & 0 & 0  \\
- 20 \left(z^{2} + z + 4\right) & - 4 \left(9 z^{2} + 3 z + 61\right) & - 10 \left(z^{2} + z + 4\right) & - 60 \left(z^{2} + z + 4\right) & 4 z \left(13 z + 20\right) & 12 \left(5 z^{2} + 8 z + 24\right) & 0 & 0  \\
0 & 4 & 0 & 0 & 0 & 0 & -12 & 12  \\
0 & -4 & 0 & 0 & 0 & 0 & 12 & -12  \\
\rmatrix \nn \\ \nn \\
B^{(0)}_3 &=& \lmatrix{cccccccc}
12 \left(z^{2} + 2 z + 5\right) & 0 & 4 \left(z^{2} + 4 z + 7\right) & - 4 \left(5 z^{2} + 2 z + 17\right) & - 12 z \left(z + 5\right) & 20 \left(z^{2} + z + 4\right) & 0 & 0  \\
0 & 0 & 0 & 0 & 0 & 0 & 0 & 0  \\
4 \left(z^{2} + 4 z + 7\right) & 0 & 2 \left(z^{2} + 4 z + 7\right) & - 4 \left(z^{2} + 4 z + 7\right) & - 28 z & 4 \left(z^{2} + 7 z - 21\right) & 4 & -4  \\
- 4 \left(5 z^{2} + 2 z + 17\right) & 0 & - 4 \left(z^{2} + 4 z + 7\right) & 12 \left(5 z^{2} + 2 z + 17\right) & 4 z \left(13 z + 17\right) & - 60 \left(z^{2} + z + 4\right) & 0 & 0  \\
- 12 z \left(z + 5\right) & 0 & - 28 z & 4 z \left(13 z + 17\right) & 60 z^{2} & - 4 z \left(13 z + 20\right) & 0 & 0  \\
20 \left(z^{2} + z + 4\right) & 0 & 4 \left(z^{2} + 7 z - 21\right) & - 60 \left(z^{2} + z + 4\right) & - 4 z \left(13 z + 20\right) & 12 \left(5 z^{2} + 8 z + 24\right) & 0 & 0  \\
0 & 0 & 4 & 0 & 0 & 0 & -12 & -12  \\
0 & 0 & -4 & 0 & 0 & 0 & -12 & -12  \\
\rmatrix \\ \nn \\
B^{(0)}_4 &=& \lmatrix{cccccccc}
- \left(z - 1\right)^{2} & 0 & 0 & z^{2} + 22 z + 25 & - z \left(z - 1\right) & - z^{2} - 19 z - 4 & 0 & 0  \\
0 & 0 & 0 & 0 & 0 & 0 & 0 & 0  \\
0 & 0 & 0 & 0 & 0 & 0 & 0 & 0  \\
z^{2} + 22 z + 25 & 0 & 0 & - z^{2} + 50 z + 47 & - z \left(3 z + 25\right) & \left(z - 32\right) \left(z + 3\right) & 4 & -4  \\
- z \left(z - 1\right) & 0 & 0 & - z \left(3 z + 25\right) & - z^{2} & - z \left(z - 4\right) & 0 & 0  \\
- z^{2} - 19 z - 4 & 0 & 0 & \left(z - 32\right) \left(z + 3\right) & - z \left(z - 4\right) & - z^{2} + 8 z - 24 & 0 & 0  \\
0 & 0 & 0 & 4 & 0 & 0 & 1 & -3  \\
0 & 0 & 0 & -4 & 0 & 0 & -3 & 1  \\
\rmatrix \nn
\eea

\bea
\label{bbresult4}
B^{(1)}_1 &=& \lmatrix{cccccccc}
- 48 \left(7 z^{2} + 4 z + 25\right) & 24 \left(z^{2} + z + 4\right) & 12 \left(4 z^{2} + z + 13\right) & 0 & 8 z \left(41 z + 87\right) & 0 & -16 & 0  \\
24 \left(z^{2} + z + 4\right) & 0 & 0 & 0 & - 12 z \left(z + 5\right) & 0 & 0 & 0  \\
12 \left(4 z^{2} + z + 13\right) & 0 & 0 & 0 & - 6 z \left(7 z + 11\right) & 0 & 0 & 0  \\
0 & 0 & 0 & 0 & 0 & 0 & 0 & 0  \\
8 z \left(41 z + 87\right) & - 12 z \left(z + 5\right) & - 6 z \left(7 z + 11\right) & 0 & - 480 z^{2} & 0 & 0 & 0  \\
0 & 0 & 0 & 0 & 0 & 0 & 0 & 0  \\
-16 & 0 & 0 & 0 & 0 & 0 & 96 & 0  \\
0 & 0 & 0 & 0 & 0 & 0 & 0 & 0  \\
\rmatrix \nn \\ \nn \\
B^{(1)}_2 &=& \lmatrix{cccccccc}
- 32 \left(5 z^{2} + 8 z + 23\right) & - 32 \left(z^{2} + z + 4\right) & 24 \left(z^{2} + 2 z + 5\right) & 0 & 32 z \left(4 z + 23\right) & 0 & 0 & 0  \\
- 32 \left(z^{2} + z + 4\right) & 4 \left(z^{2} + 10 z + 13\right) & 4 \left(7 z^{2} - 2 z + 19\right) & 0 & 128 z \left(z + 1\right) & 0 & -16 & 0  \\
24 \left(z^{2} + 2 z + 5\right) & 4 \left(7 z^{2} - 2 z + 19\right) & 2 \left(7 z^{2} + 4 z + 25\right) & 0 & - 24 z \left(2 z + 5\right) & 0 & 0 & 0  \\
0 & 0 & 0 & 0 & 0 & 0 & 0 & 0  \\
32 z \left(4 z + 23\right) & 128 z \left(z + 1\right) & - 24 z \left(2 z + 5\right) & 0 & - 736 z^{2} & 0 & 0 & 0  \\
0 & 0 & 0 & 0 & 0 & 0 & 0 & 0  \\
0 & -16 & 0 & 0 & 0 & 0 & 160 & 0  \\
0 & 0 & 0 & 0 & 0 & 0 & 0 & 0  \\
\rmatrix \nn \\ \nn \\
B^{(1)}_3 &=& \lmatrix{cccccccc}
64 \left(z^{2} + z + 4\right) & - 16 \left(z - 1\right)^{2} & - 48 \left(z^{2} + 2 z + 5\right) & 0 & - 32 z \left(z + 8\right) & 0 & 0 & 0  \\
- 16 \left(z - 1\right)^{2} & 8 \left(z - 1\right)^{2} & 4 \left(5 z^{2} + 8 z + 23\right) & 0 & 16 z \left(2 z + 1\right) & 0 & 0 & 0  \\
- 48 \left(z^{2} + 2 z + 5\right) & 4 \left(5 z^{2} + 8 z + 23\right) & 4 \left(10 z^{2} + z + 31\right) & 0 & 8 z \left(11 z + 30\right) & 0 & -16 & 0  \\
0 & 0 & 0 & 0 & 0 & 0 & 0 & 0  \\
- 32 z \left(z + 8\right) & 16 z \left(2 z + 1\right) & 8 z \left(11 z + 30\right) & 0 & 256 z^{2} & 0 & 0 & 0  \\
0 & 0 & 0 & 0 & 0 & 0 & 0 & 0  \\
0 & 0 & -16 & 0 & 0 & 0 & -64 & 0  \\
0 & 0 & 0 & 0 & 0 & 0 & 0 & 0  \\
\rmatrix \\ \nn \\
B^{(1)}_4 &=& \lmatrix{cccccccc}
32 \left(z^{2} + z + 4\right) & - 4 \left(z - 1\right)^{2} & - 8 \left(z^{2} + z + 4\right) & - 24 \left(z^{2} + 4 z + 7\right) & - 16 z \left(z + 8\right) & 24 \left(z + 2\right)^{2} & 0 & 0  \\
- 4 \left(z - 1\right)^{2} & 0 & 0 & 12 \left(z^{2} + 4 z + 7\right) & 4 z \left(2 z + 1\right) & - 12 \left(z + 2\right)^{2} & 0 & 0  \\
- 8 \left(z^{2} + z + 4\right) & 0 & 0 & 12 \left(2 z^{2} - z + 5\right) & 2 z \left(11 z + 16\right) & - 6 \left(4 z^{2} + z + 16\right) & 0 & 0  \\
- 24 \left(z^{2} + 4 z + 7\right) & 12 \left(z^{2} + 4 z + 7\right) & 12 \left(2 z^{2} - z + 5\right) & 0 & 8 z \left(8 z + 21\right) & 0 & -16 & 0  \\
- 16 z \left(z + 8\right) & 4 z \left(2 z + 1\right) & 2 z \left(11 z + 16\right) & 8 z \left(8 z + 21\right) & 128 z^{2} & - 48 z \left(z + 2\right) & 0 & 0  \\
24 \left(z + 2\right)^{2} & - 12 \left(z + 2\right)^{2} & - 6 \left(4 z^{2} + z + 16\right) & 0 & - 48 z \left(z + 2\right) & 0 & 0 & 0  \\
0 & 0 & 0 & -16 & 0 & 0 & -32 & 0  \\
0 & 0 & 0 & 0 & 0 & 0 & 0 & 0  \\
\rmatrix \nn 
\eea
\end{landscape}

\section*{Acknowledgments}

I would like to thank Sandor Katz, Tamas Kovacs, Yu Nakayama and Slava Rychkov for very useful and enlightening
discussions. The work was in part supported by the Hungarian National Research, Development and Innovation
Office (NKFIH) grant KKP126769.

\end{document}